\documentclass[epj]{svjour}
\usepackage{epsfig} 
\usepackage{times}
\begin{document}
%
%
\title{Aspects of {\boldmath $\phi$}-meson production in 
proton-proton collisions}
\author{A.~Sibirtsev\inst{1}, J.~Haidenbauer\inst{2}
and U.-G.~Mei{\ss}ner\inst{1,2}} 
\institute{Helmholtz-Institut f\"ur Strahlen- und Kernphysik (Theorie), 
Universit\"at Bonn, Nu\ss allee 14-16, D-53115 Bonn, Germany \and
Institut f\"ur Kernphysik (Theorie), Forschungszentrum J\"ulich GmbH,
D-52425 J\"ulich, Germany}
\date{Received: date / Revised version: date}

\abstract{We analyze near-threshold cross section data for the
reaction $pp{\to}pp\phi$ published by the DISTO collaboration
and recent, still preliminary results presented by the ANKE
Collaboration. We formulate a procedure to evaluate the
OZI ratio at low energies by taking into account corrections from 
the kinematics and the final-state interaction. Combining the new data with 
the few measurements available at higher energies we give a limit for
the OZI rule violation and estimate the possible contribution from
a five-quark baryonic resonance coupled to the $\phi{p}$ system.
}  


\authorrunning{A.~Sibirtsev, J.~Haidenbauer, U.-G.~Mei{\ss}ner}
\titlerunning{${\rm\phi}$-meson production in proton-proton collisions}

\maketitle

\section{Introduction}
\label{sec:intro}
$\phi$-meson production in hadronic reactions motivates
both experimental and theoretical activities for several reasons. 
A well known argument is the possible violation of the Okubo-Zweig-Iizuka rule 
\cite{Okubo,Zweig,Iizuka}
and the role of the strangeness content of the nucleon. Moreover, $\phi{p}$ 
production allows to search for cryptoexotic baryons with 
hidden strangeness.

The OZI rule states that the
production of open as well as hidden strangeness from any initial
state that does not contain a $s{\bar s}$ component is strongly suppressed
relative to the production of non-strange states. Therefore, studies
of the OZI rule are generally based on considering measurements of
corresponding $\phi$-meson and $\omega$-meson production reactions. 
The cross section ratio, $R_{\phi{/}\omega}$, is then compared with
the limit proposed by Lipkin~\cite{Lipkin}. The latter follows from 
SU(3) symmetry and involves the experimentally known deviation
$\Delta\theta_V{=}3.7^o$ from the ideal mixing angle 
between singlet and octet vector mesons, i.e. 
\begin{eqnarray}
R_{\phi{/}\omega}{=}\frac{g^2_{\phi\rho\pi}}{g^2_{\omega\rho\pi}}{=}
\frac{g^2_{\phi{NN}}}  {g^2_{\omega{NN}}}{=}
\frac{\sigma (\pi N{\to}\phi X)}{\sigma (\pi N{\to}\omega X)} {=}
\frac{\sigma (NN{\to}\phi X)}{\sigma (NN{\to}\omega X)} \nonumber \\
{=}\tan^2(\Delta\theta_V){=}4.2{\cdot}10^{-3},
\label{ozi1}
\end{eqnarray}
where the $g$'s denote corresponding coupling constants and 
$X$ stands for any inclusive final state that does not contain 
strange quarks. Actually Eq.~(\ref{ozi1}) is already violated 
on the level of the
$\phi\rho\pi$ and $\omega\rho\pi$ coupling constants that can be
extracted from $\phi{\to}\rho\pi$, $\omega{\to}3\pi$,
$\omega{\to}\pi\gamma$ and $\rho{\to}\pi\gamma$ 
decays~\cite{Gellmann,Meissner1}. These decays~\cite{PDG} result 
in an average ratio $R_{\phi{/}\omega} = (12.5 \pm 3.4) \cdot 10^{-3}$. 
Note that is also established~\cite{Jain} that the
$\phi{\to}\rho\pi$ decay alone violates OZI rule. 

A systematic analysis~\cite{Sibirtsev1} of experiments on $\phi$ and 
$\omega$ production in $\pi{N}$ and $NN$ reactions led to a ratio of
$R_{\phi{/}\omega} = (13.4 \pm 3.2) \cdot 10^{-3}$.  Apparently, 
all phenomenological models~\cite{Nakayama1,Nakayama2,Titov1,Titov2,Kaptari} 
that directly incorporate the 
$\phi\rho\pi$ and $\omega\rho\pi$ vertices would be able to reproduce
the large ratio given above. The same is also true for $NN\to NN\phi/\omega$
calculations~\cite{Griffiths,Sibirtsev2} which utilize 
$\pi{N}{\to}\phi{N}$ and $\pi{N}{\to}\omega{N}$ transition 
amplitudes fitted to the data. Therefore, values around 
$R_{\phi{/}\omega} {\simeq} 13{\cdot}10^{-3}$ \cite{Sibirtsev1} are
not necessarily related to the strangeness content
of the nucleon. Only ratios significantly larger than this value 
might be considered as a possible indication 
of a $s{\bar s}$ component of the initial state.

A very recent analysis~\cite{Sibirtsev3} of $\phi$ and $\omega$
photoproduction from the proton shows that the ratio  
$R_{\phi{/}\omega} = 0.8{\pm}0.2$ at photon energies above 30~GeV.
This was interpreted in terms of quark--anti-quark fluctuations of the
photon. Considering the $u{\bar u}$, $d{\bar d}$, $s{\bar s}$, $c{\bar
c}$,  $b{\bar b}$ and $t{\bar t}$ photon structure  one might
expect that at high energies, {i.e.} in the perturbative QCD regime, 
the ratios of different vector mesons approach unity, up to corrections due to
the  hadronic wave functions.  
The analysis~\cite{Sibirtsev4,Sibirtsev5} of
$\omega$ and $J/\Psi$ photoproduction results in the ratio
$R_{[J/\Psi]{/}\omega}{=}0.04{\pm}0.01$ at photon energy around
$3{\cdot}10^3$ GeV, {\it i.e.} at the  maximal energy where $\omega$-meson
photoproduction data are available. We concluded that the large 
$\phi{/}\omega$ and $[J/\Psi]{/}\omega$ ratios observed in 
photoproduction at high energies are related to the nature of the
photon itself but are irrelevant for the strangeness content of the 
nucleon. It was also noted that the $\phi$ 
photoproduction at low photon energies and for large four-momentum
transfer squared $t$ could not be understood in terms of perturbative QCD
and probably the large ratio $R_{\phi{/}\omega}{\simeq}0.1$ at $|t|{>}2$~GeV$^2$
can be only explained if one assumes a large ratio of the $\phi{NN}$ and
$\omega{NN}$ couplings. Modern dispersion analyses~\cite{Hammer,Mergell} 
of the nucleon electromagnetic form factors show that the squared ratio 
of these coupling constant is about ${\simeq}0.23$ (provided one assumes that
the whole strength in the spectral function at $t \simeq 1\,$GeV$^2$ is
entirely given by the $\phi$-meson, therefore such values should be considered
as upper limits). 

There are still some phenomena~\cite{Nomokonov} which could 
be a signal of a $s{\bar s}$ component in the nucleon. For example,
proton--antiproton annihilation at rest 
results in $R_{\phi{/}\omega}{=}0.294{\pm}0.097$ for the $\phi\gamma$
and $\omega\gamma$ final states while a ratio 
$R_{\phi{/}\omega}{=}0.106{\pm}0.012$ was found for the $\phi\pi$
and $\omega\pi$ channels. At the same time the available data for the
annihilation in flight yield a ratio 
$R_{\phi{/}\omega}$ = $(14.55{\pm}1.92){\cdot}10^{-3}$, 
which is
compatible with the ratios extracted from $\pi{N}$ and $NN$ reactions
and from vector meson decays. Again a substantial violation of the OZI 
rule was detected at low energies.

These observations in the $\gamma{p}$ and ${\bar p}p$ reactions
provide a strong motivation to further search for substantial OZI
rule violations at low energies. In 1998, the DISTO Collaboration 
reported a low energy $pp \to ppV$ measurement indicating that
the ratio $R_{\phi{/}\omega}$ is enhanced by about an order of
magnitude relative to the OZI limit given in Eq.~(\ref{ozi1}). 
The experiment was done at a proton beam energy of $T_{lab}\simeq$2.85 GeV,
{\it i.e.} at an excess energy $\epsilon{\simeq}$82~MeV above the 
$pp{\to}pp\phi$ reaction threshold. At such low energies corrections 
due to the differences in the $pp{\to}pp\phi$ and $pp{\to}pp\omega$
phase space and the $pp$ final state interaction (FSI) should be
implemented for a meaningful analysis with regard to an OZI rule 
violation. Indeed Eq.~(\ref{ozi1}) holds only for high energies and 
when the cross sections are practically energy independent. 

Since the corrections mentioned above were not fully implemented
in the analysis of Ref.~\cite{Balestra} the results given by the 
DISTO collaboration can be only considered as qualitative estimations. 
Here we describe a procedure for the
data evaluation that allows for an analysis of any 3-body final 
state. Following that procedure we extract the ratio $R_{\phi{/}\omega}$ 
from the DISTO results and from preliminary $pp{\to}pp\phi$ 
cross section data reported recently by the ANKE
Collaboration~\cite{Hartmann}. Note that the evaluation of 
the ratio $R_{\phi{/}\omega}$ requires data for
$pp{\to}pp\omega$ as well. High-statistics $\omega$ data
were published recently by the TOF Collaboration~\cite{TOF1,TOF2}.
Those data allow us to extract the $pp{\to}pp\omega$ reaction
amplitude, which can be considered an added bonus of our present study.

The manuscript is organized as follows. In Sec.~\ref{sec:method} we
present and discuss our method to extract the OZI ratio from data 
on vector meson production in proton-proton collisions. It is applied
to data from Saclay and COSY and the results are presented in 
Sec.~\ref{sec:res}. Sec.~\ref{sec:penta} is devoted to the possible
coupling of exotic states to the $\phi$-proton final state. We
end with a summary and outlook in Sec.~\ref{sec:sum}.

\section{Method for extracting the OZI ratio from proton-proton collisions}
\label{sec:method}

\begin{figure}[t]
\vspace*{-6mm}
\centerline{\hspace*{5mm}\psfig{file=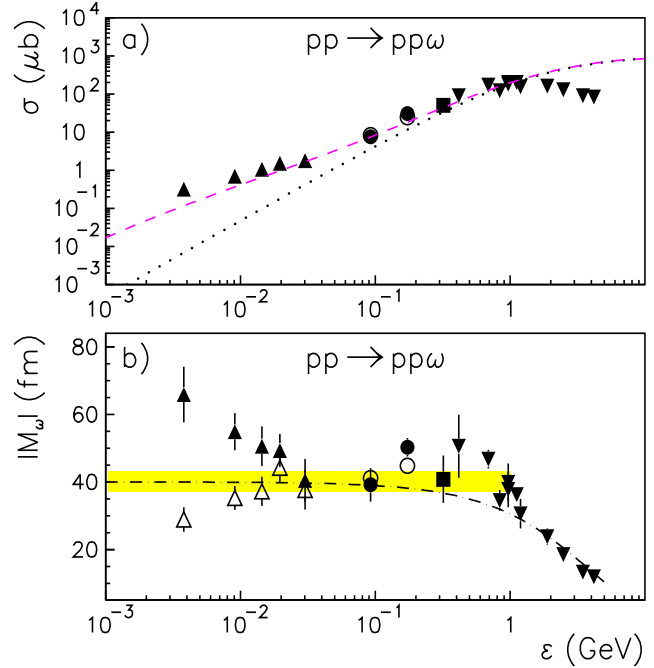,width=9.5cm,height=10.cm}}
\vspace*{-4mm}
\caption{(a) The $pp{\to}pp\omega$ cross section and (b) the average reaction
amplitude $|{\cal M}_\omega|^2$ as a function of the excess energy. 
The cross section data are from Refs.~\cite{Flaminio} 
(solid inverse triangles), \cite{Balestra} (solid squares),
\cite{Hibou} (solid triangles), \cite{TOF1} (solid circles),
and \cite{TOF2} (open circles), respectively. 
The corresponding values in (b) are extracted from the cross section data 
via Eq.~(\ref{proc1}). The open triangles correspond to the data of 
Ref.~\cite{Hibou} but using Eq.~(\ref{proc2}).
The dashed line
corresponds to Eq.~(\ref{proc1}) with $|{\cal M}|{=}40$~fm, while the 
dotted line indicates the results without FSI corrections. 
The hatched area indicates the average amplitude used in the present
paper for the OZI analysis. The dash-dotted line is a phenomenological fit 
to the data of the form ${\cal M}{=}40~{\rm fm}\,\exp(-0.27\epsilon)$.
}
\label{phima1}
\end{figure}

Using non-relativistic 3-body phase space and applying corrections due
to the FSI between the protons, the average $pp{\to}pp\omega$ reaction
amplitude $|{\cal M}|^2$ can be extracted from the reaction cross section
$\sigma$ by means of~\cite{Sibirtsev6}
\begin{eqnarray}
\sigma(\epsilon)=\frac{\sqrt{m_N^2m_\omega}}{2^7\pi^2(2m_N+m_\omega  )^{3/2}}
\, \, \frac{\epsilon^2}{\sqrt{s^2-4sm_N^2}} \nonumber \\
\times \left[ 1+\frac{4\beta^2-4\alpha^2}
{-\alpha+\sqrt{\alpha^2+m_N\epsilon}}\right] |{\cal M}|^2,
\label{proc1}
\end{eqnarray}
where $\epsilon{=}\sqrt{s}{-}2m_N{-}m_\omega$,
$\sqrt{s}$ is the invariant collision energy given in terms of the proton 
beam kinetic energy $T_{lab}$ by $s = 2m_N(2m_N{+}T_{lab})$ and $m_N$, $m_\omega$ are the
nucleon and $\omega$-meson masses, respectively. The FSI effects are taken
into account by means of the Jost function method with 
\begin{eqnarray}
|J(q)|^{-1}{=}\frac{q{+}i\beta}{q{-}i\alpha}{=}
\left[\frac{r\beta^2}{2}{+}
\frac{rq^2}{2}\right]
\left[-\frac{1}{a}+\frac{rq^2}{2}-iq\right]^{-1},
\label{josta}
\end{eqnarray}
which at small $q$ goes over into the Watson-Migdal parameterization 
and at large $q$ approaches unity. In Eq.~(\ref{josta}) $\alpha$ 
and $\beta$ are the parameters that specify the FSI between the 
protons; they are related to the scattering
parameters by 
\begin{eqnarray}
a{=}\frac{\alpha+\beta}{\alpha\beta}, \,\,\,\,\,\,
r{=}\frac{2}{\alpha+\beta}
\label{albet}
\end{eqnarray}
with $\alpha{<}0$ and $\beta{>}0$. In the present study we use
the values $\alpha = -20.5\,$MeV/c and $\beta  = 166.7\,$MeV/c. 
Integrating the square of the Jost function (\ref{josta}) over 
the nonrelativistic 3-body phase space yields the
factor in front of $|{\cal M}|^2$ in Eq. (\ref{proc1}). 

The
presently available data for the $pp{\to}pp\omega$ reaction cross
section~\cite{Flaminio,Balestra,TOF1,TOF2,Hibou} are shown in
Fig.~\ref{phima1}a) as a function of the excess energy. 
Applying Eq.~(\ref{proc1}) we can extract the average reaction amplitude 
$|{\cal M}|$. Corresponding results are displayed in Fig.~\ref{phima1}b). 
To evaluate the $\phi$/$\omega$ ratio for the DISTO and ANKE data we need
the $pp{\to}pp\omega$ reaction amplitude for excess energies in the range 
10${<}\epsilon{<}$100~MeV. The only data available in this energy range
are those of the TOF~\cite{TOF1,TOF2} and SPES-III~\cite{Hibou} collaborations. 
The reaction amplitude extracted from the SPES-III data via Eq.~(\ref{proc1})
exhibits a strong energy dependence near threshold 
which might be due to (neglecting) the finite width of the $\omega$ meson. 
Indeed one should account for the $\omega$ width,
$\Gamma$=8.49~MeV, when analyzing the data at low excess energies,
{\it i.e.} at $\epsilon{\simeq}\Gamma$. In that
case the relation between the $pp{\to}pp\omega$ cross section and
reaction amplitude is given as
\begin{eqnarray}
\sigma(\epsilon){=}\frac{1}{2^8\pi^3s\sqrt{s^2{-}4sm_N^2}}\!\!\!
\int\limits_{2m_\pi}^{\sqrt{s}-2m_N}\!\!\!
\frac{dx}{2\pi}\frac{\Gamma \, \, |{\cal M}|^2 }
{(x{-}m_\omega)^2{+}\Gamma^2/4} \nonumber \\
\!\!\!\!\!\int\limits_{4m_N^2}^{(\sqrt{s}-x)^2}\!\!\!\!\!\!\!\!dy
\frac{\sqrt{y^2{-}4ym_N^2}\sqrt{(s{-}y{-}x)^2{-}4yx^2}}{y}
\frac{y^2{-}4m_N^2{+}4\beta^2}{y^2{-}4m_N^2{+}4\alpha^2}\, .
\label{proc2}
\end{eqnarray}
Replacing the $\omega$-meson spectral distribution by the
$\delta$-function in Eq.~(\ref{proc2}) one recovers Eq.~(\ref{proc1}) 
in the non-relativistic limit. Indeed the relativistic
corrections play only a minor role at $\epsilon{<}10$~GeV, which can be
checked by numerical integration of Eq.~(\ref{proc2}). The
$pp{\to}pp\omega$ reaction amplitude extracted from SPES-III data utilizing
Eq.~(\ref{proc2}) is shown by the open triangles in
Fig.~\ref{phima1}b). Although again the amplitude deviates from a
constant at small $\epsilon$ we interpolate it as
$|{\cal M}| = (40{\pm}3)$~fm for our further OZI analysis. Finally the dashed
line in Fig.~\ref{phima1}a) shows the result of  Eq.~(\ref{proc1})
with $|{\cal M}|{=}40$~fm, while the dotted line indicates the results
without FSI corrections, {\it i.e.} with $\alpha{=}\beta$ in
Eq.~(\ref{proc1}). 

We would like to make a short comment concerning the FSI corrections, or
more precisely concerning the normalization of the lines in Fig.~\ref{phima1}a). 
Since we choose $|{\cal M}|{=}40$~fm we reasonably describe the data for 
$\epsilon{\le}1$~GeV. But the use of a constant matrix element in 
Eq.~(\ref{proc1}) apparently leads to an overestimation of the cross section 
at high energies. However, one should keep in mind that any  
phenomenological model for that reaction will 
contain form factors that depend on the squared four-momentum
transfer from the initial to the final proton. This $t$-dependence becomes
substantial~\cite{Sibirtsev6} at $\epsilon > 1$~GeV and effectively reduces
the value of the integral over the phase space. This in turn allows to describe 
data at high energies within such models~\cite{Griffiths,Sibirtsev2}.

\section{Results}
\label{sec:res}
\begin{figure}[t]
\vspace*{-6mm}
\centerline{\hspace*{5mm}\psfig{file=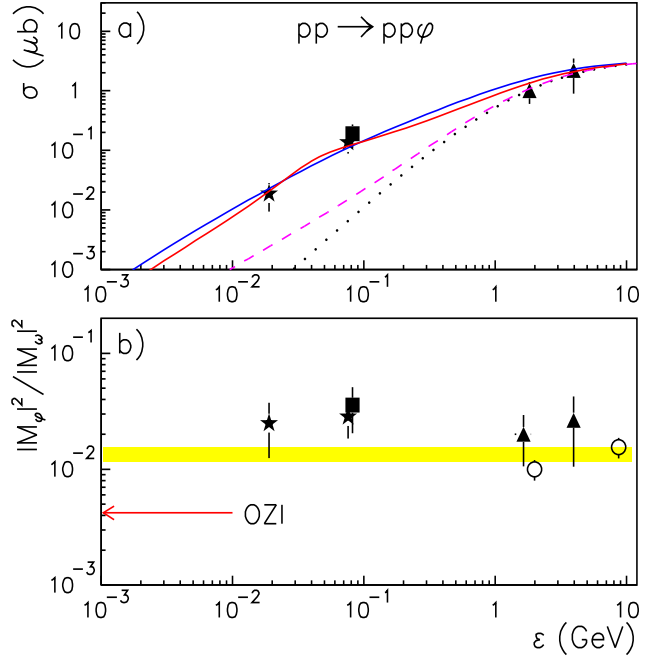,width=9.5cm,height=10.cm}}
\vspace*{-4mm}
\caption{(a) The $pp{\to}pp\phi$ cross section and (b) the ratio of
the amplitudes $|{\cal M}_\phi|^2$/$|{\cal M}_\omega|^2$ 
as a function of the corresponding excess energy. 
The square is the result from DISTO~\cite{Balestra}, the stars are the 
preliminary ANKE data~\cite{Hartmann}, while the triangles are from
Refs.~\cite{Flaminio,Blobel}. 
The open circles in (b) are from ratio measurements in the forward 
hemisphere~\cite{Arenton,Golovkin}. 
The dashed
line shows the result of Eq.~(\ref{proc1}) with $|{\cal M}| = 2.3$~fm, while 
the dotted line indicates the results without FSI corrections. 
The solid lines are the results obtained with the inclusion of a
baryonic resonance coupled to $\phi{p}$ system, discussed in Sect. 4. 
The hatched
area indicates the ratio extracted previously~\cite{Sibirtsev1} from
a combined $\pi{N}$ and $NN$ OZI analysis. The arrow indicates the 
OZI limit given in Eq.~(\ref{ozi1}).}
\label{phima1a}
\end{figure}

The formalism for extracting the reaction amplitude $|{\cal M}|^2$ for
$\phi$-meson production is the same as described in the preceeding 
section, except that now the $\phi$ mass, $m_\phi$, appears in 
Eq.~(\ref{proc1}) and the excess energy is
defined by $\epsilon{=}\sqrt{s}{-}2m_N{-}m_\phi$.
Fig.~\ref{phima1a} shows the $pp{\to}pp\phi$ cross section 
and the ratio of the $\phi$ and $\omega$ production amplitude,
$|{\cal M}_\phi|^2$/$|{\cal M}_\omega|^2$, as a function of the excess energy. 
In general, there are no measurements available
for $\omega$ as well as $\phi$ production at exactly the same
$\epsilon$ and, in addition, the $\omega$-meson data shown in
Fig.~\ref{phima1} indicate some fluctuations, which might be related
to systematics. Thus, for calculating the ratio for 
the DISTO and ANKE data we use the average amplitude 
$|{\cal M}| = (40{\pm}3)$~fm for $\omega$ meson production. The
triangles in Fig.~\ref{phima1a}b) show the ratios obtained from the data on 
the total cross sections of Refs.~\cite{Flaminio,Blobel}. The open circles 
are ratios published in Refs.~\cite{Arenton,Golovkin} which correspond
to cross section measurements in the forward hemisphere. Note that at such
high energies differences in the phase space are practically negligible.

The hatched band in Fig.~\ref{phima1a}b) indicates the ratio 
$R_{\phi{/}\omega} = $ $(13.4{\pm}3.2){\cdot}10^{-3}$ extracted from
a combined analysis of $\pi{N}$ and $NN$ data in Ref.~\cite{Sibirtsev1}, 
which is close to the value
$R_{\phi{/}\omega} = (12.5{\pm}3.4){\cdot}10^{-3}$ obtained from the
vector meson decays. The DISTO measurement results in the ratio
$R_{\phi{/}\omega}{=}(36{\pm}16){\cdot}10^{-3}$. The still preliminary 
ANKE data~\cite{Hartmann} are consistent with the DISTO result. Thus, 
the central value of the ratio is indeed a factor of 4 larger than
what was found in the other analyses. Unfortunately, however, 
the systematic uncertainties of the data on the $\phi$/$\omega$ ratio from
$pp$ collisions are rather large and, therefore, impede the claim for
a clear and more substantial OZI rule violation as compared 
to the limits found, for instance, in vector meson decays. 

\begin{figure}[b]
\vspace*{-6mm}
\centerline{\hspace*{7mm}\psfig{file=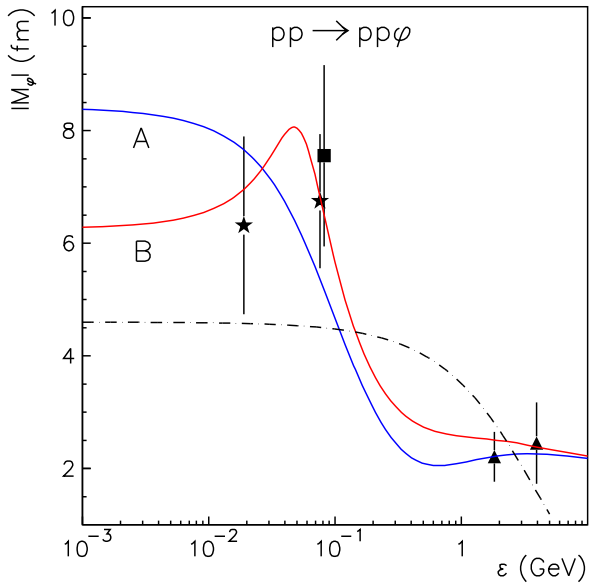,width=8.9cm,height=8.9cm}}
\vspace*{-4mm}
\caption{The $pp \to pp\phi$ reaction amplitude, $|{\cal M}_\phi|$,
as a function of the excess energy $\epsilon$. The square is
the result extracted from the DISTO data~\cite{Balestra}, 
the stars are those for the preliminary ANKE data~\cite{Hartmann}, while 
the triangles are from Refs.~\cite{Flaminio,Blobel}. The lines show results 
of a calculation that includes a baryonic resonance in the $\phi{p}$ system 
with the parameters sets A and B given by Eq.~(\ref{models}). The dash-dotted
line indicates the energy dependence of the $pp \to pp\omega$ amplitude
and is taken over from Fig.~\ref{phima1}.}
\label{phima1b}
\end{figure}

In any case, it is clear that the near-threshold cross section data
for the reaction $pp \to pp\phi$ are enhanced as compared to the 
measurements at $\epsilon > $1~GeV. This is demonstrated by 
the dashed line in Fig.~\ref{phima1a}a). It corresponds to a 
calculation using Eq.~(\ref{proc1}) and utilizing a constant reaction
amplitude, $|{\cal M}|$ = 2.3~fm, which is normalized to the 
high energy data.
In order to demonstrate that more clearly let us now use 
Eq.~(\ref{proc1}) to extract the reaction amplitude $|{\cal M}|$
from the measured cross sections. Corresponding results are
shown in Fig.~\ref{phima1b}. To facilitate a comparision with 
the energy dependence of the amplitude for the reaction 
$pp \to pp\omega$ we include here a fit to that amplitude
(dash-dotted line), taken over from Fig.~\ref{phima1}.
Given the few data available and the large error bars there is
some freedom in how the curve for $\omega$ production is superimposed
on the results for the $\phi$ production amplitude. Thus, at
present one cannot rule out that the $\epsilon$-dependence of the 
$pp \to pp\phi$ amplitude is basically the same as the one 
obtained for the reaction $pp \to pp\omega$. 
On the other hand, the data do suggest that, in the range
$0.1 < \epsilon < 1$~GeV, the $pp \to pp\phi$ amplitude 
exhibits a significantly stronger variation with energy than what 
is seen for the $\omega$ production. 
This characteristic energy dependence could be a sign
for an additional reaction mechanism in the $\phi$ production
reaction, and specifically it might be caused by the
excitation of a resonance in the $\phi{p}$ system. 
We will elaborate on this point in the next section.

\section{Speculations on exotic baryons}
\label{sec:penta}
A few years ago Landsberg proposed~\cite{Landsberg1,Landsberg2} that 
$\phi{p}$ production, which is OZI suppressed for non-resonant
reactions, is well suited for the search of cryptoexotic baryons with 
hidden strangeness, $B_\phi{=}udds{\bar s}$. 
It is expected that these
pentaquark baryons have a narrow width and decay preferentially into the
$\phi{N}$, $K{\bar K}N$ or $YK$ channels, where $Y$ stands for ground-state
or excited hyperons. Note that these decays are OZI allowed. 
Experimental
limits for the $B_\phi$ candidates  were reported in 
Refs.~\cite{Arenton,Aleev,Dorofeev1,Dorofeev2,Balatz,Antipov}. There are two
independent observations of a narrow peak in the $\Sigma(1385)^0K^+$ spectra, 
the first one with a mass $M = 2050{\pm}6$ MeV and width
$\Gamma \le50{\pm}19$ MeV \cite{Dorofeev2} and the second one with
$M = 1956^{+8}_{-6}$ MeV and $\Gamma{=}27 \pm 15$ MeV \cite{Aleev}.
The high-statistics study of Ref.~\cite{Antipov} of the $\Sigma^0K^+$ mass spectrum
indicates two exotic states with $M = 1807{\pm}7$ MeV,
$\Gamma = 62{\pm}19$ MeV and $M = 1986{\pm}6$ MeV,
$\Gamma = 91{\pm}20$ MeV. 

\begin{figure}[t]
\vspace*{-6mm}
\centerline{\hspace*{-1mm}\psfig{file=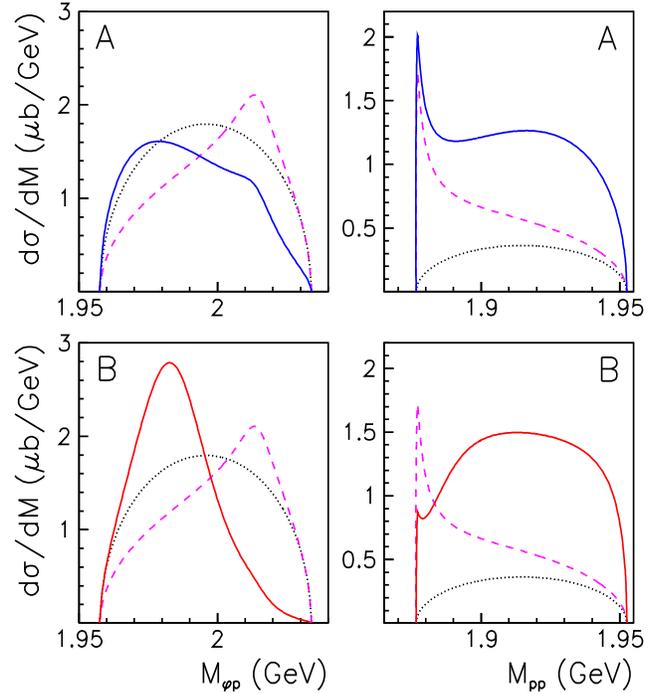,width=9.9cm,height=10.5cm}}
\vspace*{-4mm}
\caption{The $\phi{p}$ and $pp$ invariant mass spectra for the reaction
$pp \to pp\phi$ at $\epsilon = $76~MeV. The dotted lines show
the phase space distributions, the dashed lines are calculations with inclusion
of the $pp$ FSI and the solid lines are results obtained with including the $pp$ FSI 
and a $B_\phi$ pentaquark resonance with parameter sets A and B 
from Eq.~(\ref{models}).}
\label{phima1c}
\end{figure}

If cryptoexotic $B_\phi$ baryons indeed exist one expects them to 
contribute to the $pp \to pp\phi$ reaction. That was the 
basic idea of the experiment of Ref.~\cite{Arenton} where an
upper limit for $B_{\phi} \to \phi{p}$ was provided but the low
statistics did not allow to draw definite conclusions from the $\phi{p}$
mass spectrum. Note that these measurements were done at
$\epsilon \simeq 2$~GeV, where the non-resonant contribution to the 
process $pp \to pp\phi$  might dominate, as can be estimated from
Fig.~\ref{phima1b}. In that sense a study of a possible $B_\phi$ contribution
at low excess energies has advantages and we expect that COSY is
well suited for such an investigation \cite{Hartmann1}. 

To estimate the possible effect from a $B_\phi$ state, we introduce a baryonic
resonance in the $\phi{p}$ subsystem and parameterize the reaction
amplitude by 
\begin{eqnarray}
{\cal M}={\cal M}_0+\frac{c_0 e^{i\phi} M\Gamma}{M^2-s_{\phi{p}}-iM\Gamma},
\end{eqnarray}
where $M$ and $\Gamma$ are the resonance mass and width, respectively,
while $s_{\phi{p}}$ is the squared invariant mass of the $\phi{p}$
subsystem. Here we take ${\cal M}_0 = 3$~fm, in line with the data at
higher energies, cf. Fig.~\ref{phima1b}, and adjust the real
constant $c_0$ so that the $pp \to pp\phi$ data are reproduced over the
whole considered energy range. The possible phase $\phi$ between the two 
amplitudes was chosen to be $\phi$=0 for convenience (note also that the
existing data do not allow to pin down this phase). 
The curves in
Fig.~\ref{phima1b} correspond to  calculations with two different parameter
sets:
\begin{eqnarray}
&{\rm A:}& \,\, M=1956~{\rm MeV},\,\, \Gamma=120~{\rm MeV},\,\, c_0=8~{\rm fm}
\nonumber \\
&{\rm B:}& \,\, M=2000~{\rm MeV},\,\, \Gamma=\phantom{1}40~{\rm MeV},\,\,  c_0=8~{\rm fm}
\label{models}
\end{eqnarray}
Obviously it is not possible to determine the 
$B_\phi$ parameters unambiguously from the ANKE
and the DISTO data on the reaction cross section. 
Therefore, we present results from two different
sets which describe the data almost equally well. A reasonable
choice of $B_\phi$ allows to reproduce the data on $pp \to pp\phi$
cross section, as illustrated by the two solid lines in
Fig.~\ref{phima1a}a). It is clear that, in principle, the large $pp \to pp\phi$
cross section at low energies could be explained by a $B_\phi$
excitation. 

However, this is rather speculative and 
further detailed studies are necessary before conclusions
can be drawn.
First it is important to have the final results from the ANKE 
Collaboration~\cite{Hartmann1} 
with high statistical accuracy and also at low excess energies,
where the different $B_\phi$ resonances yield different energy
dependences of the  $pp \to pp\phi$ reaction cross section. It would be
also important to get data for the $pp \to pp\phi$ reaction at higher
energies to verify the strong energy dependence of the reaction amplitude
in the region $0.1 < \epsilon < $2~GeV, and specifically to map out the
energy dependence in that region in detail. This could e.g. 
be done at the JINR Nuclotron~\cite{Salmin}. 

Independently of that, a direct investigation of a possible $B_\phi$ 
contribution can be done through the analysis of the
$\phi{p}$ and $pp$ invariant mass spectra. Fig.~\ref{phima1c} shows 
results for these mass spectra for the $pp \to pp\phi$ reaction at
$\epsilon = 76$~MeV, {\it i.e.} for the current experimental setup of
ANKE. The dotted lines represent the phase space distribution with $|{\cal
M}| = {\rm const.}$, while the dashed lines are calculations where the $pp$
FSI is taken into account via Eq.~(\ref{josta}). The solid lines in
Fig.~\ref{phima1c} are predictions that include the $pp$ FSI and a $B_\phi$
resonance with parameter sets A and B given in Eq.~(\ref{models}).
It is clear that the identification of a $B_\phi$ contribution requires high
statistical accuracy and a high mass resolution. These are prerequisites
for any experiment with the aim to resolve the issue of the possible existence 
of exotic pentaquark baryons with hidden strangeness.

\section{Summary and outlook}
\label{sec:sum}

In this paper, we have studied aspects of vector meson production in
proton-proton collisions. We have proposed a strategy that allows
to extract the OZI ratio $R_{\phi/\omega}$ correcting for the
kinematical differences in the processes $pp\to pp\omega$ and $pp \to
pp\phi$, respectively, and for the final-state interactions. The
resulting values collected in Fig.~\ref{phima1a} are somewhat larger but 
still compatible with the extraction of $R_{\phi/\omega}$ from 
$\pi N$ and $NN$ reactions. We have also pointed out 
a strong energy dependence of the $pp \to pp\phi$ reaction amplitude
for excess energies $0.1 < \epsilon < 1\,$GeV. This might be indicative
of a resonance coupling strongly to the $\phi p$ system. We have presented
results assuming the presence of a cryptoexotic baryon with 
hidden strangeness. However, to establish such a state, a precise 
measurement of invariant mass distributions, see Fig.~\ref{phima1c}, 
is mandatory. Such measurements could be performed 
with the ANKE detector at COSY.

\subsection*{Acknowledgements}
We would like to thank K.-Th.~Brinkmann, \,  H.-W.~Hammer, 
C.~Hanhart, M.~Hartmann, Y.~Maeda,
J.~Ritman and  E.~Roderburg for useful
discussions.  This work was partially  supported  by Deutsche
Forschungsgemeinschaft  through funds provided to the SFB/TR 16
``Subnuclear Structure of Matter''. This research is part of the \, EU
Integrated \, Infrastructure \, Initiative Hadron Physics Project under
contract  number RII3-CT-2004-506078. A.S. acknowledges support by the
COSY FFE grant No. 41760632 (COSY-085).

\end{document}